\documentclass[twoside,twocolumn,a4paper]{article}
\usepackage{graphicx}
\usepackage{blindtext}
\usepackage{lettrine}

\usepackage[sc]{mathpazo}
\usepackage[T1]{fontenc} 
\usepackage{microtype} 

\usepackage[english]{babel} 

\usepackage[hmarginratio=1:1,top=20mm,columnsep=15pt,left=18mm]{geometry} 
\usepackage[hang, small,labelfont=bf,up,textfont=it,up]{caption} 
\usepackage{booktabs} 

\usepackage{lettrine} 

\usepackage{enumitem} 
\setlist[itemize]{noitemsep} 

\usepackage{abstract}

\usepackage{titlesec} 
\titleformat{\section}[block]{\large\scshape\centering}{\thesection.}{1em}{} 
\titleformat{\subsection}[block]{\large}{\thesubsection.}{1em}{} 

\usepackage{fancyhdr} 
\usepackage{titling} 
\usepackage{hyperref} 
\usepackage{authblk}

\pretitle{\begin{center}\Huge\bfseries} 
\posttitle{\end{center}} 
\title{Crossover from ballistic to diffusive vortex motion in convection} 
\author[1]{Kai Leong Chong}
\author[2]{Jun-Qiang Shi}
\author[2]{Shanshan Ding}
\author[1]{Guang-Yu Ding}
\author[2]{Hao-Yuan Lu}
\author[2]{Jin-Qiang Zhong}
\author[1,3]{Ke-Qing Xia}
\affil[1]{Department of Physics, The Chinese University of Hong Kong, Shatin, Hong Kong, China.}
\affil[2]{School of Physics Science and Engineering, Tongji University, Shanghai 200092, China.}
\affil[3]{Department of Mechanics and Aerospace Engineering, Southern University of Science and Technology, Shenzhen, Guangdong 518055, China.}
\date{\today}
\renewcommand{\maketitlehookd}{%
\begin{abstract}
\noindent 
Vortices play an unique role in heat and momentum transports in astro- and geo-physics, and it is also the origin of the Earth's dynamo. A question existing for a long time is whether the movement of vortices can be predicted or understood based on their historical data. Here we use both the experiments and numerical simulations to demonstrate some generic features of vortex motion and distribution. It can be found that the vortex movement can be described on the framework of Brownian particles where they move ballistically for the time shorter than some critical timescales, and then move diffusively. Traditionally, the inertia of vortex has often been neglected when one accounts for their motion, our results imply that vortices actually have inertial-induced memory such that their short term movement can be predicted. Extending to astro- and geo-physics, the critical timescales of transition are in the order of minutes for vortices in atmosphere and ocean, in which this inertial effect may often be neglected compared to other steering sources. However, the timescales for vortices are considerably larger which range from days to a year. It infers the new concept that not only the external sources alone, for example the solar wind, but also the internal source, which is the vortex inertia, can contribute to the short term Earth's magnetic field variation. 
\end{abstract}
}

\begin{document}

\maketitle

\lettrine{B}{}rownian motion is an example of stochastic processes that happens widely in nature \cite{pusey2011brownian}. Date back to 1905, Einstein proposed the theoretical explanation to the movement of pollen particle within the thermal bath \cite{einstein1905molekularkinetischen}. When the inertia of the particle taken into account, it is expected that the particle should undergo ballistic motion in short time and then replaced by diffusive motion when their motion being randomized. This transition occurs in very short timescale that the direct observation can only be achieved after hundred years of research \cite{huang2011direct}. However, Brownian motion is found to exist in many other examples besides the pollen particle. For instance, the particles in membranes in biological systems \cite{saffman1975brownian}. Also for colloidal particle, it is found that the active colloid exhibits the transition from ballistic to diffusive motion as they self-propel in short time, which is in contrast to the pure diffusive motion for passive colloidal particle. For all the examples, the common belief is that the objects should have distinct density or mass difference from their environment such that inertia plays a role initially. In this article, we demonstrate by both the experiments and numerical simulations that the whirling fluid -- vortex -- also exhibits the properties of the inertial Brownian particles. Within certain critical timescale, the memory of vortex motion becomes effective such that it persists to drift along the previous direction. This results entails the capability of predicting the vortex motion within certain period of time in astro-and geo-physical systems.

\begin{figure*}[h!]
	\centering
	\setlength{\unitlength}{\textwidth}
	\includegraphics[width=1.0\unitlength]{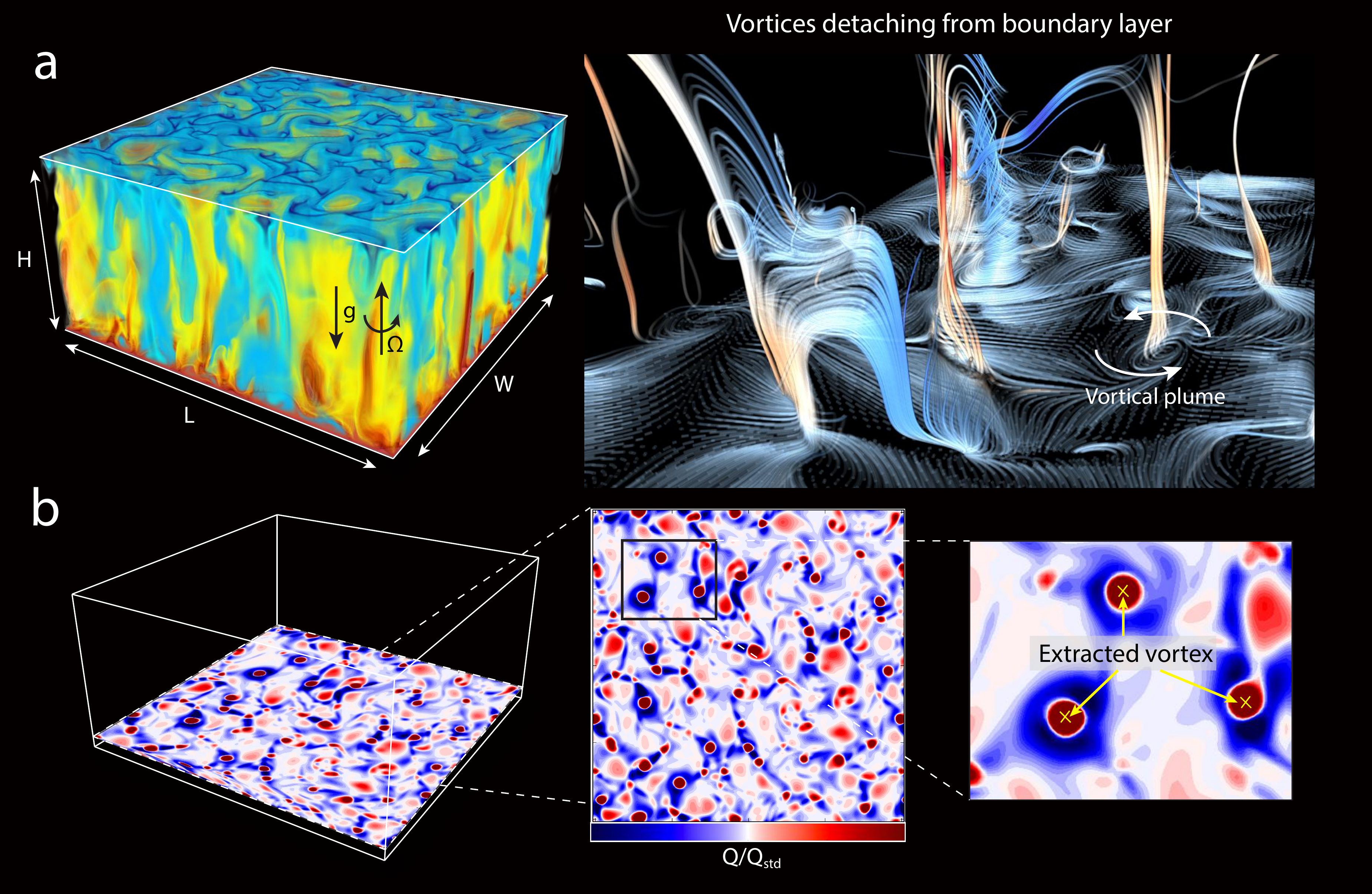}
\caption{(a) Snapshots of the temperature $\theta$ (left) and streamline originates from the lower thermal boundary layer. (b) Snapshots of $Q/Q_{std}$ (see main text for the definition of Q) taken horizontally at the edge of thermal boundary layer for $Ek=4\times10^{-5}$, $1\times10^{-5}$ at $Ra=10^8$, and a demonstration of the extracted vortex. The location of vortices center are marked as yellow crosses.}
\label{fig1}
\end{figure*}

For many occasions in astrophysics, geophysics and meteorology, the thermal convection is simultaneously influenced by rotation. The existence of Coriolis force leads to the formation of vortex \cite{hopfinger1993vortices}, which appears ubiquitously in nature for example the tropical cyclones in atmosphere \cite{emanuel2003tropical}, oceanic vortices \cite{flament1996three}, long-lived giant red spot in Jupiter \cite{marcus2004prediction}. Also, one intriguing example is the convective Taylor columns in outer core of Earth, which is believed to be playing a major role in the Earth's dynamo \cite{glatzmaiers1995three,glatzmaier1996rotation,busse1998convection,olson1999numerical,kuang1997earth,busse1998convection,christensen1999numerical,glatzmaier1999role,olson1999polar} and thus related closely to the Earth's magnetic field variation and the corresponding seismic activities \cite{varotsos1984physical,varotsos1984physical2}. A formidable question is always open in astro-and geophysical communities whether we can predict the movement of vortices within certain period of time. It therefore leads to the fundamental study on the features of vortices in laboratory. The model system to study the rotating convection is the so-called rotating Rayleigh-B\'enard (RB) convection \cite{chandrasekhar2013hydrodynamic,ahlers2009rmp,lohse2010arfm,chilla2012epj,xia2013taml,clercx2018mixing} which is a fluid layer of height $H$ heated from below and cooled from above while being rotated about the vertical axis at angular velocity $\Omega$. Here the temperature difference destabilize the flow such that the convection occurs and becomes turbulent when the thermal driving is sufficiently strong. For this model system, three dimensionless parameters can be used to characterize the flow dynamics which are  the Rayleigh number $Ra=\alpha g \Delta T/\kappa\nu$, the Prandtl number $Pr=\nu/\kappa$ and the Ekman number $Ek=\nu/2\Omega H^2$. Here $\alpha$ is the thermal expansion coefficient, $\kappa$ and $\nu$ are the rate of viscous and thermal diffusion.

\begin{figure*}[h!]
	\centering
	\setlength{\unitlength}{\textwidth}
	\includegraphics[width=1.0\unitlength]{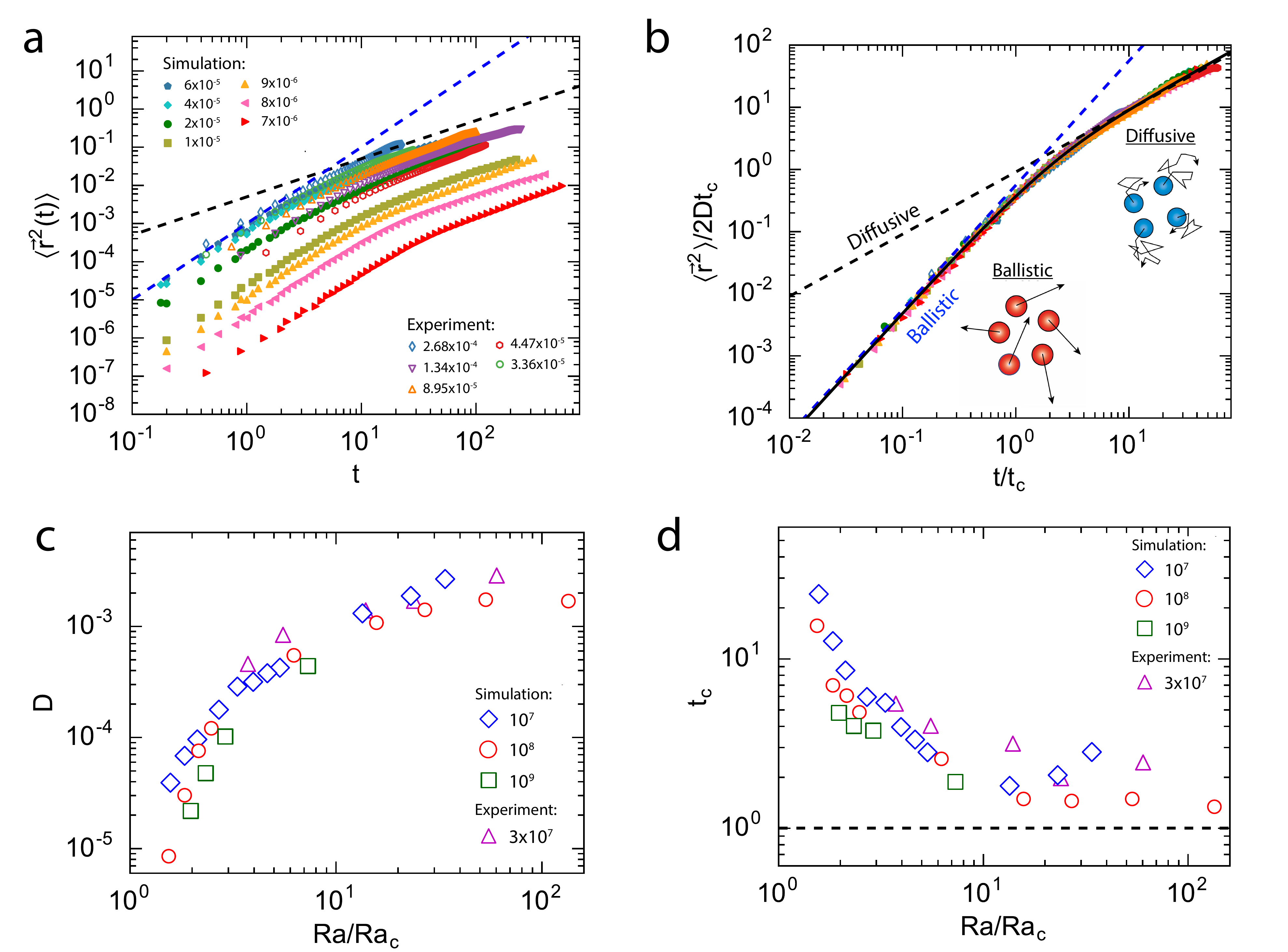}
\caption{(a) The mean square displacement (MSD) of the vortices as a funtion of time, and (b) Normalized MSD as a function of $t/t_c$. Solid symbols refer to numerical results of $Ra=1\times10^8$, while open symbols refer to experimental results of $Ra=3\times10^7$. (c) Diffusion coefficient of vortices $D$ and (d) the characteristic timescale for motion transition $t_c$ as a function $Ra/Ra_c$.}
\label{fig2}
\end{figure*}

Rotation introduces richness of structures. The general feature of rotating convection flow is as such: In the case without rotation, it is observed that fragmented thermal plumes detach from the thermal boundary layer and transport to the opposite boundary layer. However when the rotation plays a role, vortical structure becomes widely present from where we can observe the fluid parcels transport spirally up/down as shown in Fig. \ref{fig1}. It is also known that those vortical plumes resulting from Ekman pumping are favourable to heat transport \cite{zhong2009prl,Chong2015prl}. When the rotation rate becomes rapid yet not too strong to completely laminarize the flow, the Taylor-Proudman effect \cite{proudman1916motion,taylor1923stability} becomes dominant which suppresses the flow variation along the axis of rotation. The resultant flow field is the long-lived columnar structure extending throughout the entire cell height which is so-called the convective Talyor columns \cite{boubnov1986experimental,grooms2010prl}. Those vortices are not steady but move around as they interact with the background turbulent flow. 


\begin{figure*}[h!]
	\centering
	\setlength{\unitlength}{\textwidth}
	\includegraphics[width=1.0\unitlength]{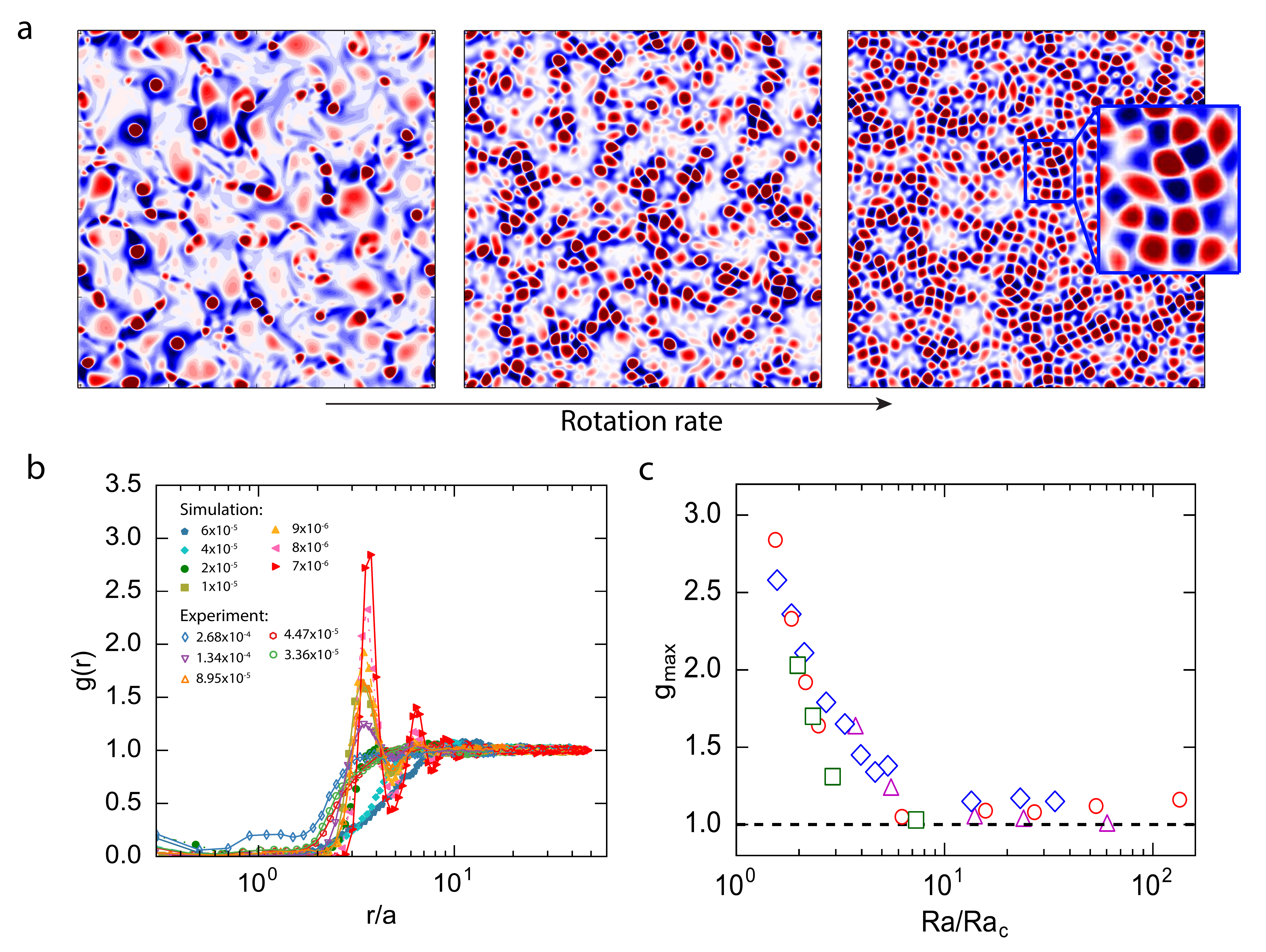}
\caption{(a) Snapshots of $Q/Q_{std}$ (see main text for the definition of Q) taken horizontally at the edge of thermal boundary layer for $Ek=4\times10^{-5}$, $1\times10^{-5}$ and $7\times10^{-6}$ at $Ra=10^8$. (b) Radial distribution function $g(r)$ as a function of $r/a$, where $a$ is the average radius of vortices. (c) The maximum radial distribution function $g_{max}$ versus $Ra/Ra_c$.}
\label{fig3}
\end{figure*}

This study contains both the experiments and the numerical simulations. For experiments, $Ra$ is fixed at $3\times10^7$ while $Ek$ is varied from $3.36\times10^{-5}$ to $2.68\times10^{-4}$. For numerical simulations, $Ra$ spans from $10^7$ to $10^9$ while $Ek$ ranges from $3.36\times10^{-5}$ to $2.68\times10^{-4}$. In order to minimize the influence of sidewalls in experiments, the large diamenter-to-height aspect ratio of about $4$ is adopted. Also, periodic boundary condition is adopted in numerical simulation with width-to-height aspect ratio of $2$. Besides, in the present study we only consider the influence by Coriolis force but neglecting the effect of centripetal force which is valid in experiment for small enough Froude number (usually for $Fr \ll 0.05$ \cite{zhong2009prl}). Next, in order to compare experimental and numerical results, all the physical parameters are made dimensionless by using the buoyancy timescale (also known as the free-fall timescale), top-and-bottom temperature difference and the system height. To study the motion and distribution of vortices, those structures should be extracted in advance, which can be done by the well-known extraction method so-called the Q-criterion. Figure \ref{fig1}b demonstrates a typical field of Q quantity and the examples of extracted vortices. For the details of the extractions, we refer to the Materials and Methods. 

\section{Vortices horizontal motion}
First we examine the motion of vortices by tracking the individual vortex from a sequence of snapshots sampling smaller than $1$ free fall time unit such that the movement of vortex is smooth in this time frame. With the trajectories of vortices, the behaviour of vortex motion can be characterized by using mean square displacement (MSD) which is calculated as  $\langle \vec{r}^2(t) \rangle = \frac{1}{N}\sum^N_{i=1}(\vec{r}_i(\tau+t)-\vec{r}_i(\tau))^2$ where $N$ is the total number of trajectories. Figure \ref{fig2}a shows the MSD versus time $t$ for $Ra=10^8$ for simulation data and for $Ra=3\times10^7$ for experimental data with different $Ek$ where three order of magnitude in timescales is covered for each $Ek$. Surprisingly, the MSDs for different $Ek$ and $Ra$ are found to have the universal trend, and the scaling of MSD $\langle \vec{r}^2(t) \rangle \sim t^{\alpha}$ exhibits multiple regimes. For short time, the vortex exhibited ballistic motion with $\alpha=2$. Eventually, the vortex motion undergoes transition to normal diffusion with $\alpha=1$. It is interesting to observe this transition from ballistic to diffusive motion that resembles the behaviour of Brownian particles in a thermal bath, especially the existence of ballistic motion because the motion of vortices were previously thought as over-damped, meaning a negligibly small inertia compared to the viscous damping. Using the approach for Brownian particle, the behaviour of vortex motion can then be described by the solution of a Langevin equation given below:

\begin{figure*}[h!]
	\centering
	\setlength{\unitlength}{\textwidth}
	\includegraphics[width=1.0\unitlength]{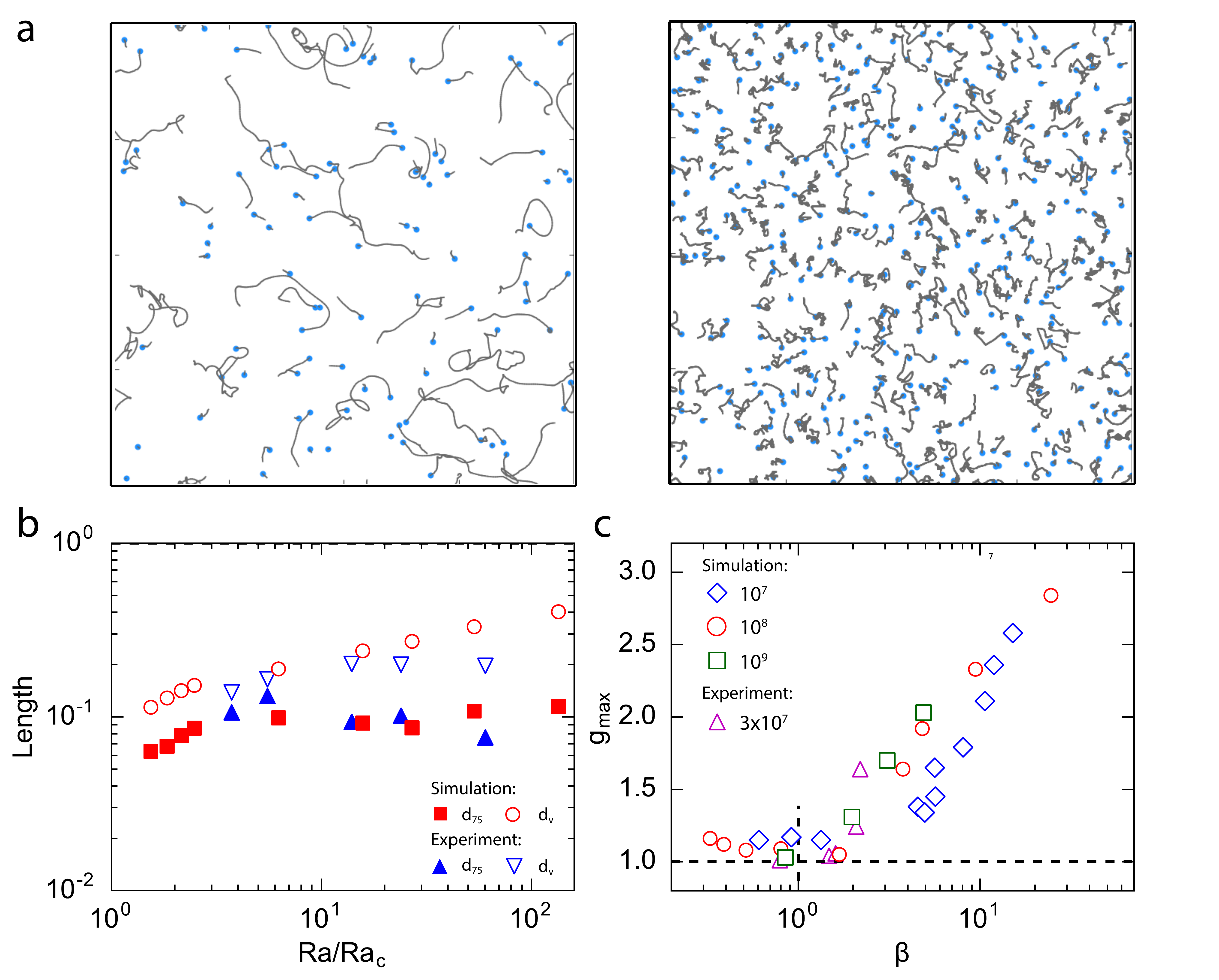}
\caption{(a) Trajectories of vortices for $Ek=1\times10^{-4}$ (left) and $Ek=7\times10^{-6}$ (right) at $Ra=1\times10^8$, indicated by the grey curves. The blue dots refer to the end of these trajectories. (b) The average distance between vortices (open symbols) and 75 percentile of the distance that vortices can reach (solid symbols), as a function of $Ra/Ra_c$ for $Ra=1\times10^8$ (simulation, red symbols) and $Ra=3\times10^7$ (experiment, blue symbols) (c) The maximum radial distribution function $g_{max}$ versus $\beta$, which is the ratio between the Brownian timescale and the relaxation timescale.}
\label{fig4}
\end{figure*}

\begin{equation}
\ddot{\vec{r}}=-\dot{\vec{r}}/t_c+\vec{\xi }(t),
\end{equation}
\begin{equation}
\left \langle \vec{\xi }(t) \right \rangle=0,
\end{equation}
\begin{equation}
\left \langle \vec{\xi }(t') \cdot \vec{\xi }(t'') \right \rangle=\frac{2D}{t_c^2}\delta (t'-t'')
\end{equation}

where $\vec{\xi }(t)$ is the noise term and $\delta$ is the Dirac delta function. And $t_c$ is the characteristic timescale for motion transition and $D$ is the diffusion coefficient of vortices in thermal turbulence. The MSD can be evaluated from the Langevin equation which yields

\begin{equation}
\label{msdeq}
\frac{\left \langle \vec{r}^2(t) \right \rangle}{2Dt_c}=\frac{t}{t_c}(1-\frac{t_c}{t}(1-exp(-\frac{t}{t_c})))
\end{equation}

With the measured MSD for each $Ek$ and $Ra$, one can fit the equation (\ref{msdeq}) to the data to obtain $D$ and $t_c$ for each case. Next by plotting $\left \langle \vec{r}^2(t) \right \rangle/2Dt_c$ versus $t/t_c$ one finds that all the measured MSDs collapse nicely onto a single curve $f(x)=x(1-\frac{1}{x}(1-exp(-x)))$ which again confirms that the motion of the individual vortex have the same dynamical behaviour regardless of $Ra$ and $Ek$. The Langevin equation contains two fitting parameters $D$ and $t_c$ where they vary with different control parameters. However, it is remarkable to observe that when $D$ and $t_c$ are plotted against $Ra/Ra_c$ (where $Ra_c$ is the critical Rayleigh number given by $Ra_c=8.7Ek^{-4/3}$ \cite{chandrasekhar2013hydrodynamic}), the data collapse more or less onto a single trend as shown in Fig. \ref{fig2}c,d. It suggests that the rescaled $Ra$ actually be the proper parameter here to describe the dynamics of vortex motion. It is noteworthy that $t_c$ approaches to one when $Ra \geq 10Ra_c$. As the timescale is normalized by the buoyancy timescale,  it suggests that the buoyancy time becomes the dominant scale for controlling this transition when $Ra$ is larger than about $10Ra_c$.



\section{Vortices distribution}
Besides the dynamical behaviour of vortices, the distribution of vortices is also examined. In Fig. \ref{fig3}a, the snapshots of $Q/Q_{std}$ at the edge of thermal boundary layer are demonstrated for $Ek=4\times10^{-5}$, $1\times10^{-5}$ and $7\times10^{-6}$ at $Ra=10^8$. When $Ek$ varies from $4\times10^{-5}$ to $7\times10^{-6}$, i.e., increasing the rotation rate, several changes in the vortex distribution can be distinguished. First, the number of vortices increases with the rotation rate such that the dilute distribution of vortices become highly concentrated. Second, we also observe that for sufficiently high rotation rate, the vortices tend to form the vortex-grid structure. Zooming in local region in Fig. \ref{fig3}a can clearly see that there is regular pattern for such vortex-grid structure: Vortices depicted by reddish color form square lattice and in between them, there are bluish localized area denoting the region having strong stress (definition of Q quantity).

Next we quantify the structural formation by using the radial distribution function $g(r)$. It is often used in condense matter for characterizing the distribution of particle pairs. The function $g(r)$ is defined as the ratio of the actual number of vortex lying between $r$ and $r+\Delta r$  to the expected number for uniform distribution. With this definition, $g(r)$ equals to one for the case of randomly distributed vortices. Figure \ref{fig3}b shows $g(r)$ versus the distance between vortices $r$ normalized by the average radius of vortices $a$ ($a$ is evaluated from the average area of vortex with assuming the area as perfect circle). Taking results from numerical simulations as an example. For any $Ek$, when it is close to the center of vortex, the value of $g(r)$ is less than one because a single vortex has its own occupying area. For $Ek>10^{-5}$, it is still probable to have the neighbouring vortex appears for distance smaller than vortex radius since vortex splitting occurs frequently. However, for $Ek\leq10^{-5}$, when the convective Taylor columns are completely formed, those columnar vortex acts like a rigid columns such that $g(r)$ begins with value very close to zero. Another distinguishing feature is the occurrence of peaks for $g(r)$. For $Ek>10^{-5}$, $g(r)$ will gradually saturate at the value of one, representing the random distribution for the neighbouring vortices. In contrast, for $Ek\leq10^{-5}$, $g(r)$ overshoots the value of one before eventually saturating to one, and this pronounced peak is an indication of forming vortex-grid structure. For the smallest $Ek$($=7\times10^{-6}$, $g(r)$ even exhibit multiple peaks which is an evidence on vortex-grid. It is also noteworthy that the peak location all occurs at about $r=4a$ because the vortices are not closely packed but separated by a localized region of strong stress as shown qualitatively in Fig. \ref{fig3}a. The experimental results also agree nicely with the numerical results. We further extract the maximum of $g(r)$ and plot it against the rescaled $Ra$, it is interesting to observe that data points for different $Ra$ and $Ek$ collapse onto a single curve. For $Ra\geq10Ra_c$, $g_{max}$ is close to one, indicating the random distribution of vortices. In contrast, for $Ra<10Ra_c$, $g_{max}$ increases with decreasing rescaled $Ra$.

\section{Discussions}
The above results lead to the question why the out-of-random vortex distribution does not be reflected from the individual vortex motion. Here by the example of vortex trajectories shown in Fig. \ref{fig4}a, it can be seen that the vortex motion is actually very localized. In Fig. \ref{fig4}b, we have further evaluated the traveled distance for 75-percentile of the vortices $d_{75}$ compared to the vortex separation $d_v$. It is clear that the majority of vortices actually move about its own colony without intervening other vortices. Next we explain what determines the spatial structure of vortices to change from a random distribution to the one with regular grid. It is actually due to the two competing dynamics where their significance can be described by the relaxation timescale and the Brownian timescale respectively. Here the relaxation timescale $t_s$ is defined as $1/\langle \parallel S \parallel \rangle_{x,y,t}$ where $\langle \parallel S \parallel \rangle_{x,y,t}$ is the magnitude of normal stress averaging over time and over horizontal plane at the edge of thermal BL. And the Brownian timescale is defined as $a^2/D$ where $a$ is the vortex radius. The ratio of the two timescales $\beta= \langle \parallel S \parallel \rangle_{x,y,t}a^2/D$ measures the tendency to form vortex aggregation, and this ratio is somewhat similar to the P\'eclet number used in Stokesian dynamcis of colloidal dispersions \cite{brady1988stokesian}. While both become larger for stronger rotation, but their relative strength determines the morphology of vortices distribution. For $\beta \leq 1$, vortex motion is dominated by the diffusion, and thus any vortex structure induced by normal stress would be distorted by the rapid diffusion (with large $D$) and therefore, the distribution of vortices appears to be random. On the other hand, for $\beta > 1$, diffusion is too weak to break the connection between vortices and thus aggregation is resulted. In Fig. \ref{fig4}c, we plot the peak value of radial distribution function $g(r)$ versus $\beta$. Indeed, $g_{max}$ starts to increases from one when $\beta$ becomes larger than one.

So far we have demonstrated some universal features of vortex in thermal convection. An question is the implication to the the cases in astro- or geo-physics. One of the interesting results is the existence of ballistic motion for vortex and here we estimate the corresponding transition timescale for different astro- or geophysical aspects. First, we consider the situation of vortices in troposphere \footnote{Estimates on $Ra$ and $Ek$ are based on the physical parameters as such: $H\approx10^4m$, $\Delta T\approx100K$, $g\approx10ms^{-2}$, $\alpha\approx0.003K^{-1}$, $\nu\approx2\times10^{-5}m^2s^{-1}$, $\kappa\approx10^{-5}m^2s^{-1}$ and $\Omega\approx10^{-4}s^{-1}$.} with $Ra\approx10^{22}$ and $Ek\approx10^{-9}$. One can obtain that it is far from onset of convection ($Ra/Ra_c\approx10^9$), and thus we use the buoyancy timescale $\sqrt{H^4/(\nu\kappa Ra)}$ to estimates the transition time $t_c$ which gives $t_c\approx1min$. Similarly for oceanic vortices \footnote{Physical parameters $H\approx10^4m$, $\Delta T\approx10K$, $g\approx10ms^{-2}$, $\alpha\approx0.0003K^{-1}$, $\nu\approx10^{-5}m^2s^{-1}$, $\kappa\approx10^{-3}m^2s^{-1}$ and $\Omega\approx10^{-4}s^{-1}$.}, we obtain $t_c\approx10min$. Usually only the diurnal or seasonal changes of vortices will be focused, the minute-scale transition time implies that the inertial effect of vortex can actually be neglected in those cases. However using parameters to estimate the situation of Earth's liquid-metal outer core, considerably large transition timescale has been obtained. For Earth's outer core, $Ra$ is difficult to be preciously determined but it is accepted that $Ra$ should range from $10^{22}$ to $10^{30}$ and $Ek\approx10^{-15}$. By using the physical parameters $H\approx2\times10^6$, $\nu\approx10^{-6}m^2s^{-1}$ and $\kappa\approx10^{-5}m^2s^{-1}$, one can estimate $t_c$ is in the order of hour to year. In astrophysics, there is conventional thought that the short term variation (timescale of years or less) of Earth's magnetic field should primarily caused by the external sources, for example the solar winds \cite{bloxham1985secular}. Based on our estimation, the inertia of vortex should be another significant factor on columnar vortex movement in the core of Earth and thus it is an example of internal sources affecting the short term variation in the Earth's magnetic field. It also hints at the potential of forecasting the activity of Earth's magnetic field and the related seismic activities.

\section{Acknowledgements}
K.L.C., G.Y.D., and K.Q.X. were supported in part by the Hong Kong Research Grants Council under Grant No. CUHK404513 and a NSFC$/$RGC Joint Research Grant N\_CUHK437$/$15, and through a Hong Kong PhD Fellowship. J.Q.Z. was partially supported by a NSFC$/$RGC Joint Research Grant No.~11561161004.

\section{Materials and Methods}
\subsection{Experimental set-ups}
The experimental apparatus had been used for several previous investigations of turbulent rotating RBC [cite refs.]. For the present study we installed a new cylindrical sample cell that had a diameter D=240.0 mm and a height L=63.0mm, yielding an aspect ratio $\Gamma$=D/L= 0.38. Its bottom plate, made of 35mm thick oxygen-free copper, had a finely machined top surface that fit closely into a Plexiglas side wall, and was heated from below by a uniformly-distributed electric wire heater. The top plate of the cell was a 5mm thick sapphire disc that was cooled from above through circulating temperature-controlled water. For flow visualization and velocity measurement, a particle-image-velocimetry (PIV) system was installed that consists of three main components: a solid-state laser with the light-sheet optics; neutrally-buoyant particles suspended in the flow; and a CCD camera. Both the convection apparatus and the PIV system are mounted on a rotating table that operates in a range of the rotating rate 0<=$\Omega$<=6.283*0.4 rad/s. The measuring region of the velocity field presented in this work was a central square area of 164mm*136mm of the horizontal plane at a fluid height z=L/4. In each velocity map 103*86 velocity vectors were obtained with a spatial resolution of 1.6mm. For a given Ek number, we took image sequences consisting of 18000 velocity maps at time intervals of 0.5s, corresponding to an acquisition time of 2.5 hours.

\subsection{Numerical method}
We consider the Navier-Stokes equation in Cartesian coordinate with Oberbeck-Boussinesq approximation\cite{kaczorowski2013jfm,kaczorowski2014jfm}.
\begin{equation}
	\label{equ:governing_equation1}
	\nabla\cdot u=0 \label{equ:continuity}
\end{equation}
\begin{equation}
	\label{equ:governing_equation2}
	\frac{\partial u}{\partial t}+u\cdot\nabla u=-\nabla p+\nu\nabla^2u-2\Omega\times u+\theta e_z \label{equ:momentumn}
\end{equation}
\begin{equation}
	\label{equ:governing_equation3}
	\frac{\partial \theta}{\partial t}+u\cdot\nabla \theta=\kappa\nabla^2\theta \label{equ:temperature}
\end{equation}
where $u$, $p$ and $\theta=T-(T_{hot}+T_{cold})/2$ are velocity, pressure and reduced temperature respectively. $\nu$ stands for the kinetic viscosity, $\kappa$ the thermal diffusivity, and $\Omega$ the angular velocity of the convective cell. The governing equation is solved in non-dimensional form. Physical quantities in the governing equation are non-dimensionalized by $x_{ref}=H$, $u_{ref}=\left(\alpha gH\Delta\right)^{1/2}$, $T_{ref}=\Delta$ and $t_{ref}=x_{ref}/u_{ref}$, where $H$ is the height of the convective cell, $\alpha$ the thermal expansion coefficient, $g$ the gravitational acceleration and $\Delta=T_{hot}-T_{cold}$ is the temperature difference between hot and cold plane. Two dimensionless control parameters determine the physical properties of the systen: $Ra=\alpha g\Delta L^3/(\kappa\nu)$ is the Rayleigh number indicating the strength of thermal driving, and $Pr=\nu/\kappa$ is the ratio between momentum and thermal diffusivity. The geometrical properties of the convective cell is described by and the aspect ratio $\Gamma=L/H$, where $L$ is the width of the convective cell. $\Gamma$ is fixed to be 4 in all our simulations. Periodic boundary condition is applied to the sidewall. The top and bottom plates are no-slip and isothermal. Eq.\ref{equ:governing_equation3} is solved by the multiple-resolution version of CUPS \cite{Chong2018}, which is a fully parallelized direct numerical simulation (DNS) code based on finite volume method with 4th order precision. Temperature and velocity are discreted in a staggered grid. In thermal convection with $Pr>1$, the smallest length scale is the Batchelor length $\eta_b/H=\eta_k/Pr$, where $\eta_k$ is the Kolmogorov length scale $\eta_k/H=\left(Ra\epsilon_u/Pr\right)^{-1/4}$, and $\epsilon_u=\nu\sum_i\sum_j\left(\frac{\partial u_i}{\partial x_j}\right)^2$ is the dimensionless viscous dissipation. Both Batchelor and Kolmogorov length scale is spatially intermittent, depending on the local viscous dissipation. Since boundaries are no-slip in our simulation, strong shearing occurs near boundaries and induces large $\epsilon_u$  and small length scale. To resolve these flow structures with small length scale, we use structured meshes which are refined near top and bottom boundaries, and equidistant in the two horizontal directions. As $\eta_b$ is smaller than $\eta_k$, the temperature solver requires higher spatial resolution than the momentum. Thus, for single-resolution scheme, the resolution requirement is determined by the temperature solver, which induces unneccessary compution in the momentum solver. To increase computational efficiency without any sacrifice in precision, we use a multiple-resolution strategy, which means the momentum equation is solved in a coarser grid than the temperature. The grid spacing we used in our simulations resolve both the Batchelor and Kolmogorov length scales. The temporal integraton of the governing equations is carried out by an explicit Euler-leapfrog scheme, that the convective and diffusive terms are updated using the leapfrog and the Euler forward method respectively. For the details of our code please refer to the previous paper on Rayleigh-Benard convection\cite{Chong2018}.

\subsection{Extraction of vortex}
We extract vortices based on the Q-criterion \cite{hunt1988eddies} which considers the quantity Q defined by $Q=0.5(\parallel \omega \parallel^2-\parallel S \parallel^2)$ where $\omega$ is the vorticity tensor and $S$ is the rate-of-strain tensor and $\parallel A \parallel = \sqrt{Tr(AA^T)}$. Here a single vortex is defined by the connected region satisfying $Q>Q_{std}$, with $Q_{std}$ being the standard deviation of Q, which can discern the vortices from background fluctuation. The center of vortex can further be identified by the location with maximum Q.








\end{document}